\documentclass[12pt]{iopart}
\usepackage{graphicx}
\usepackage{dcolumn}
\usepackage{color}
  
\begin{document}
  
 \title
{Novel ballistic to diffusive crossover in  the dynamics of a one dimensional Ising model
with  variable range of interaction}


\author	{Soham Biswas, Parongama Sen }
{\address {Department of Physics, University of Calcutta,\\
92 Acharya Prafulla Chandra Road, Kolkata 700009, India}}
\ead{soham.physics@gmail.com; parongama@gmail.com}

 \date{\today}
\begin{abstract}

The idea that the dynamics of a  spin  
is determined by the size of its neighbouring domains was 
recently introduced (S. Biswas and P. Sen, Phys. Rev. E {\bf 80}, 027101 (2009)) in a Ising spin model 
 (henceforth, referred to as model I). 
 A parameter $p$ is now defined to modify the dynamics such that
a spin can sense domain sizes up to $R = pL/2$ in a one dimensional system 
of size $L$. For the cutoff factor $p \to 0$, the dynamics is Ising like and the   domains grow with time $t$ diffusively  as 
$ t^{1/z}$ with $z=2$, while  for $p=1$, the original model I showed
ballistic dynamics with $z \simeq 1$. For intermediate values of $p$, 
the domain growth, magnetisation and persistence show model I like  
behaviour up to a macroscopic crossover time $ t_1 \sim pL/2$.  Beyond $t_1$,
characteristic power law variations of the dynamic quantities are 
no longer observed. The total time
to reach equilibrium is found to be  $t = apL + b(1-p)^3L^2$, from which we 
conclude that the later time behaviour is diffusive. 
We also consider the case when  a random but quenched 
 value of  $p$ is used  for each spin for which  ballistic  behaviour is once again
obtained.


\end{abstract}


\section{Introduction}
Dynamical  phenomena is an important topic in statistical physics. 
Physical  quantities in self organized and/or driven systems
show rich  time dependent behaviour in many cases. Some of the 
dynamical phenomena which have attracted a lot of attention are 
critical dynamics, quenching and coarsening, reaction diffusion systems, random walks etc. 

In most of these phenomena, we find there is a single timescale leading to  uniform  time 
 dependent behaviour which  in many cases 
is a power law decay or growth \cite{bray}. However,
in some complex systems,  
it has been observed that the dynamics is governed by a distinct short
time behaviour followed by a different behavior at long times. For example, in
spin systems, at criticality, the order parameter is observed to grow 
for a macroscopically short time \cite{early} while at longer times it decays in an  expected 
power law manner. For correlated random walks, e.g., the persistent random walk
on the other hand, one finds a ballistic (i.e., when the root mean square (rms)  displacement scales linearly
with time)
to diffusive (rms displacement varying as the square root of time) crossover in the dynamics \cite{persrw}.
Random walks on small world networks show a completely opposite behaviour, the number of distinct sites
visited by the walker has an initial diffusive scaling followed by a ballistic variation  with time \cite{swrw}. This is also true for a biased random walker.

In this paper, we study a dynamical model of Ising spins in one dimension
which is governed by a single parameter.  
The system is a generalised version of a recently 
  proposed  model
in \cite{biswas-sen} (which we refer to as model I henceforth) 
where the state of the spins ($S = \pm 1$) may change in  two situations:
first when its two
neighbouring domains  have opposite polarity,  and in this case
the  spin  orients itself along the 
spins of the neighbouring domain with the  larger size.
This case may arise only when the spin  is at the boundary of the two
domains.
The neighbouring domain sizes are calculated excluding the spin itself, however, even if it is included, there is no change in the 
results.
A spin is also flipped when it is sandwiched between two domains of spins
with same sign.
 When  the two neighbouring domains of the spin are of the same size but have opposite polarity,
the spin will change its orientation with fifty percent probability.
 Except for this rare event the dynamics in the above model is deterministic. 
This dynamics leads to a  homogeneous state of either all spins
up or all spins down. Such evolution to absorbing homogeneous 
states are known to occur in systems belonging to
directed percolation (DP) processes, zero temperature Ising model, voter model etc. \cite{absorb,vote}.

Model I 
was introduced in the context 
of a social system where the binary 
opinions of individuals are 
 represented by    up and down spin states.
In opinion dynamics models, such representation of opinions by Ising or Potts
spins is quite common \cite{opinion1}. The key feature is the interaction 
of the 
individuals which may lead to phase transitions between a homogeneous state to a heterogeneous state in many cases \cite{opinion2}.

Model I showed the existence of 
novel  dynamical  behaviour  in a coarsening process when compared to the
dynamical behaviour of DP processes, voter model, Ising models etc. \cite{hinrich2,stauffer2,sanchez,shukla,derrida}. The domain sizes 
were observed to grow as $t^{1/z}$ with the  exponent $z$ very close to unity. It may be noted that
the dynamics of a domain wall can be visualised as the movement of a walker and therefore the 
value $z\simeq 1$ indicated that the effective walks are ballistic.
When   stochasticity is introduced in this model, such that spin flips 
are dictated by a  so called ``temperature'' factor, it shows a 
robust behaviour in the sense that only for  the temperature going to infinity 
there is  conventional Ising model like behaviour with $z=2$, i.e., the domain wall dynamics becomes diffusive in nature \cite{psnew}.

In this work, we have introduced the parameter $p$, which we call the cutoff factor, such that the maximum size of the neighbouring 
domains a spin can sense is given by $R = pL/2$ in a one dimensional system of $L$ spins 
with periodic boundary condition. 
It may be noted that for $p=1$, we recover the original model I where there
is effectively no restriction on the size sensitivity of the spins. $R=1$ corresponds
 to the nearest neighbour Ising model where $p\rightarrow 0$ in the thermodynamic limit.

By the introduction of the parameter $p$ we have essentially defined a
restricted neighbourhood of influence on a spin. Thus here we have a
finite neighbourhood to be considered, which is like having
a model with finite long range interaction.
In addition, here we impose the condition that within this restricted neighbourhood,
the domain structure is also important in the same way it was in 
model I.
If one considers  opinion dynamics systems (by which  model I  was originally inspired),
the domain sizes represent some kind of social pressure.
A finite cutoff (i.e., $p < 1$)
puts a restriction on the domain sizes which  may  correspond to geographical, political, cultural   boundaries etc.
The case with uniform cutoff signifies that all the individuals
have same kind of restriction;  we have also
considered the case with random cutoffs which is perhaps closer to reality.

In the next section, we describe the dynamical rule and 
quantities estimated. 
We  present the results for the case when $p$ is same for 
all spins in section III and IV and in section V we consider the
case when the values of $p$ for each spin is random, lying between
zero and unity and constant over time for each spin.
In the last section, we end with concluding remarks. 

\section{Dynamical rule and quantities calculated}

As mentioned before, only the spins at the boundary of a domain wall can  change
its state. When sandwiched between two domains of same sign, it will be always flipped. On the other hand, for other boundary spins (termed the target spins henceforth), there will be
two neighbouring domains of opposite signs. For such spins, 
 we have the following dynamical 
scheme:
let $d_{up}$ and $d_{down}$ be the sizes of the two neighbouring
domains of type up and down of a target spin  (excluding itself). In model I,
the dynamical rule was like this: 
if $d_{up}$ is greater (less) than $d_{down}$, the target spin will be up (down) and if 
 $d_{up} = d_{down}$ the target spin is flipped with probability 0.5.
Now, with the introduction of $p$,  the definition of $d_{up}$ and $d_{down}$ are
modified: $d_{up} = {\rm{min}}\{R, d_{up}\}$ and 
similarly  $d_{down} = {\rm{min}}\{R, d_{down}\}$ while the same dynamical rule applies.

As far as dynamics is concerned, we investigate 
primarily the 
time dependent behaviour of the order parameter, fraction of domain walls   and
the persistence probability. The order parameter is  given by
 $m = \frac{|L_{up} - L_{down}|}{L}$
where $L_{up}~~ (L_{down})$ is the number of up (down) spins in the system
and $L = L_{up}+ L_{down}$, the total number of spins.
This is identical to the (absolute value of) magnetisation in the Ising model.

The average fraction of domain walls  $D_w$, which is the average number of domain walls divided by the 
system size $L$ is identical to the 
inverse of average domain size. Hence the dynamical evolution of the order parameter and fraction of domain walls 
is  expected to be governed by the dynamical exponent $z$; $m \propto t^{1/(2z)}$ and 
$D_w \simeq t^{-1/z}$ \cite{bray}. 

The persistence probability $P(t)$ 
of a spin is the probability that it remains in its original state 
up to time $t$ \cite{derrida} is also estimated. $P(t)$  has been shown to have a power law decay in many systems with 
an associated exponent $\theta$.    
The persistence probability, in finite systems has been shown to obey the following scaling form \cite{pray,bcs}
\begin{equation}
P(t,L) \propto L^{-\alpha} f(t/L^z).
\label{alpha}
\end{equation}
The exponent $\alpha= \theta z$ is associated with the 
saturation value of the persistence probability at $t\to \infty$ when 
$P_{sat}(L) = P(t \to \infty, L) \propto L^{-\alpha}$ \cite{pray}. 

In the simulations, we have generated systems of size $ L \leq 6000$ with a minimum of 2000 initial 
configurations for the maximum size in general. 
Depending on the system size and time to equilibriate, maximum iteration times have been
set. Random updating process has been used to control the spin flips. In general,  the error bars in the data are less than the   
size of the data points in the figures and therefore not shown.

\section{Case with finite $R$ ($p \to 0$)}

\begin{figure} [ht]
\begin{center}
 
 \rotatebox{0}{\resizebox*{5cm}{!}{\includegraphics{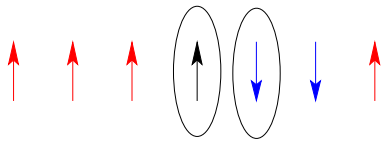}}}
\caption{ 
A schematic picture to show the dynamics in the present model for a   finite value of $R$.
Both the encircled spins will  change their state with fifty percent probability for the nearest neighbour Ising model ($R=1$). For $R=2$, the  encircled spin on the left 
will flip with probability 1/2 while  the  one on the right  will flip with probability 1. For $R=3$, the left one will not flip but the right  one will.}
\label{schematic}
\end{center}

\end{figure}

In this section, we discuss the  case when $R$ is  finite. Effectively this 
means that $R$ does not scale with $L$ and is kept a constant for all system sizes.
Since  $R$ is kept finite,  
  expressing $R=pL/2$  implies 
$p \to 0$
in the 
  the thermodynamic limit.  
For $R=1$, the model is same as the Ising model as the dynamical rule is identical to the  zero temperature Glauber dynamics.
But it may be noted that making $R>1$ will 
make the dynamical rules different from the case of $R=1$;  as an example 
we show  in Fig. \ref{schematic} how making $R=2$ or $3$  changes the 
dynamical rule compared to $R=1$.

\begin{figure} [ht]
\begin{center}
 \rotatebox{0}{\resizebox*{8.5cm}{!}{\includegraphics{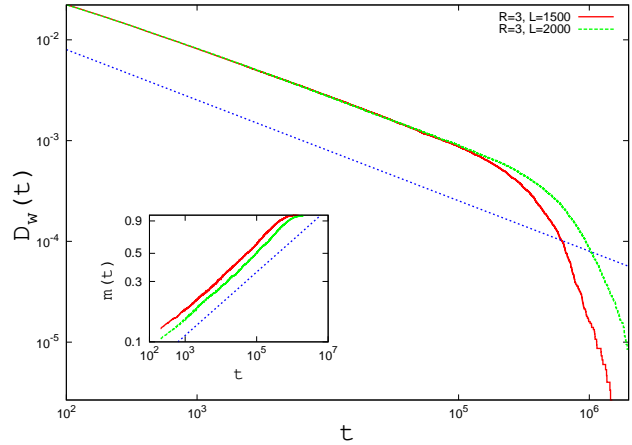}}}
\caption{
Decay of the fraction of domain walls $D_w(t)$ with time for $R=3$ and two different system sizes shown in a 
log-log plot. The dashed line has slope equal to 0.5. Inset shows growth of magnetisation $m(t)$ with time for $R=3$; the  dashed line here has slope equal to 0.25.}
\label{domainmagR3}
\end{center}
\end{figure}

\begin{figure} 
\begin{center}
 \rotatebox{270}{\resizebox*{6cm}{!}{\includegraphics{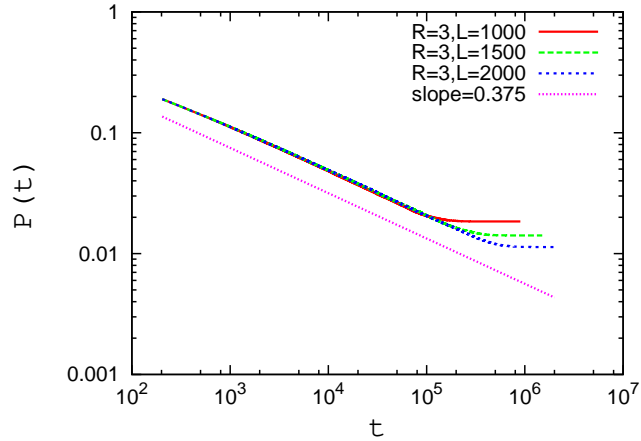}}}
\caption{
Decay of persistence probability $P(t)$ with time for three different sizes shown in a log-log plot.   The straight line has  slope 0.375.}
\label{persR3}
\end{center}
\end{figure}

We have simulated systems with $R=2$ and $R=3$ which show
that the  dynamics leads to the equilibrium configuration of all spins up/down.
Not only that, the dynamic exponents also turn out to be identical to 
those  corresponding to the nearest neighbour Ising values (i.e.,  $\theta = 0.375$ and $z = 2$). As $R$ is increased, the finite size effects become stronger, however, it is 
indicated that the Ising exponents will prevail as the system size becomes 
larger. 
In an indirect way, we have  shown later that 
$z=2$ as $p \rightarrow 0$ using a general scaling argument. 
The behaviour of the different dynamic quantities
for $R=3$ are shown in Figs \ref{domainmagR3} and \ref{persR3}.

\section{Case with  $p > 0$}

In this section, we discuss the  case when $p$ is finite. We also assume that 
$p$ is uniform,  which means each spin experiences the same cutoff.

The equilibrium behavior is same for all $p$, i.e.,  starting from a random initial configuration, 
the  dynamics again leads to a final state with $m=1$, i.e., all spins up or all spins down.
For $p=1$, that is in model I, it was numerically obtained that $\theta \simeq	 0.235$ and   $z \simeq 1.0$ giving  
$\alpha \simeq 0.235$,
while in the one dimensional Ising model $\theta = 0.375$ and $z=2.0$ (exact results) giving $\alpha =0.75$.
It is  clearly indicated that  though model I and the Ising model have  identical equilibrium behaviour, they 
belong to two different dynamical classes which correspond to $p = 1$ and
$p \to 0$  limit respectively of the present model.
It is therefore of interest to investigate the dynamics in the intermediate range of $p$. 

\subsection{Results for $0<p<1$}

\begin{figure} [ht]
\begin{center}
 \rotatebox{0}{\resizebox*{8cm}{!}{\includegraphics{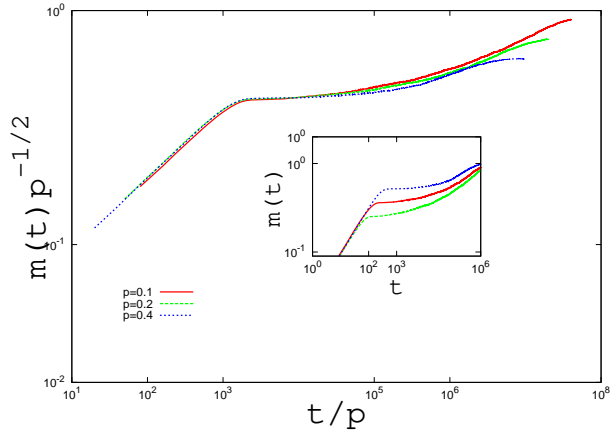}}}
\caption{ The collapse of scaled order parameter versus scaled time for different values of $p$,
shows $z=1$ for $t<t_{1}$ . Inset shows unscaled data. System size $L=3000$.}
\label{magcolp}
\end{center}
\end{figure}

Drastic changes in the dynamics  are noted for
finite values of $ p <1$. The behaviour of all the three quantities, $m(t)$, $D_w$ and $P(t)$ shows the common feature of a power law growth or decay with time up to an initial time
$t_1$
 which 
 increases with $p$. The power law behaviour is followed by a  very slow variation 
of the quantities over a much longer interval of time, before they attain the equilibrium values. The  power law behaviour in the early time occur with
exponents consistent with model I, i.e., $z\simeq1$ and $\theta \simeq 0.235$. This early time behaviour accompanied by
model I exponents is easy to explain: it occurs while the domain sizes are less than $pL/2$ such that the size sensitivity 
does not matter and the dynamics is identical to that in model I. As the domain size increase beyond this value, the sizes of the neighbouring domains as sensed by the boundary spin   
become equal making the dynamics stochastic rather than deterministic as a result of which the dynamics becomes much slower.

\begin{figure} [ht]
\begin{center}
 \rotatebox{0}{\resizebox*{8cm}{!}{\includegraphics{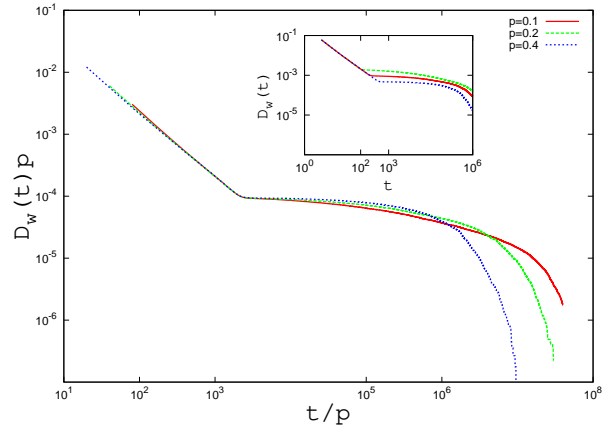}}}
\caption{The collapse of scaled fraction of domain walls versus scaled time for different values of $p$;
shows $z=1$ for $t<t_{1}$. Inset shows unscaled data. System size $L=3000$. }
\label{domaincolp}
\end{center}
\end{figure}

\begin{figure} [ht]
\begin{center}
 \rotatebox{0}{\resizebox*{8cm}{!}{\includegraphics{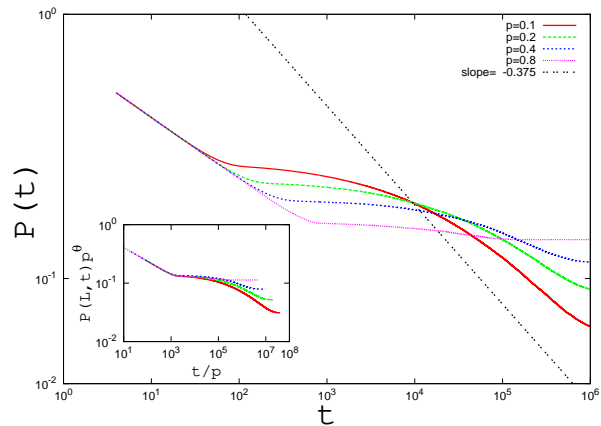}}}
\caption{ 
Persistence probability  versus time  for different values of $p$; the straight line with slope 0.375 shown for comparison.
Power law behaviour can be observed only at the initial time. 
 System size $L=3000$.
Inset shows the collapse of scaled persistence probability versus scaled time 
indicating  $z=1$ for $t<t_{1}$. 
}
\label{percolp}
\end{center}
\end{figure}

We thus argue that since domain size $ \sim t^{1/z}$, the time up to which 
model I behaviour will be observed is $t_1= (pL/2)^{z}$. Since $z$ for model I is $1$ we expect that $t_1= pL/2$. 
For a fixed size $L$ one can then consider the  scaled time variable 
$t^\prime = t/p$, and plot the relevant scaled quantities against $t^\prime$  for different 
values of $p$ to  get a data  collapse up to $t_{1}^\prime = t_1/p$, 
independent of $p$. We indeed observe this, in Figures \ref{magcolp}, \ref{domaincolp} and  \ref{percolp}, the scaling plots as well
as the raw data are shown. From the raw data, $t_1$ is clearly seen to be  different for different $p$.

Although the model I behaviour is confirmed up to $t_1$ and explained easily, beyond $t_1$, the raw data 
do not give any information about the dynamical exponents $z$ and $\theta$ as no straight forward power law fittings are possible. While an alternative method to calculate $\theta$ is not known,  one may have 
an estimate of $z$ using an indirect method. It has been shown recently that for  various dynamical Ising models,
the  time $t_{sat}$  to reach saturation varies as $L^x$ where $x$ is identical to the dynamical exponent $z$ \cite{psnew,pssdg}.
One may attempt to do the same here.

\begin{figure} [ht]
\begin{center}
 \rotatebox{270}{\resizebox*{5.6cm}{!}{\includegraphics{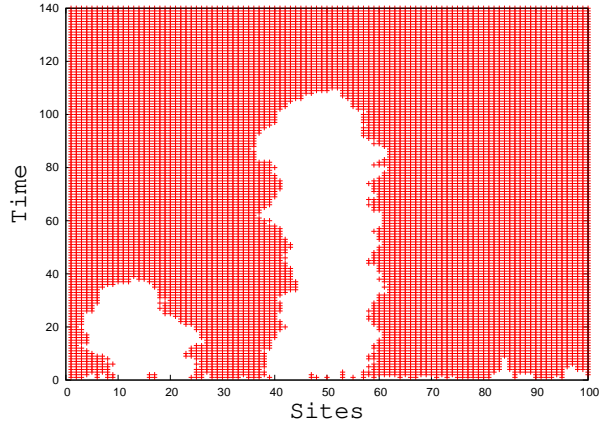}}}
\caption{ Snapshot for $p<1.0$ ($p=0.4$) for system size $L=100$.}
\label{snap}
\end{center}
\end{figure}

Actually it  is possible to find out theoretically the form of $t_{sat}$ from the qualitative behaviour of the dynamical quantities described above
and the snapshot of the system (Fig. \ref{snap}) at times beyond $t_1$. At $t > t_1$,  the domain sizes of the neighbours of any spin at the boundary appear equal
such that the domain walls perform random walks slowing down the annihilation process. Domain walls annihilate only after  one of the neighbouring domains
shrinks to a size $< pL/2$ again.  
In a small system, one can see that the slow process 
continues with  only two domain walls separating two domains 
remaining  in the system at later times (Fig. \ref{snap}). 
Even in larger systems, there will be only a few  domain walls 
remaining making 
 $D_{w} ~ \propto 1/N$ at $t > t_1$ as we note from  the inset of  Fig \ref{domaincolp}:
 $D_{w}$  remains close to $O(1/N)$ for a long time before 
going to  zero.

Thus $t_{sat}$ will have two components, $t_1$, already defined and $t_2$, the time during which there is a slow
variation of quantities over time and the last two domains remain. While $t_1 \propto pL$, one can argue that $t_2\propto (1-p)^3 L^2$.
The argument runs as follows: Let us for convenience consider the open boundary case. Here, the 
size sensitivity of the spins is $R^{open} = qL$ where $0 \leq q  \leq 1$ with the system assuming the model
I behaviour for $ q \geq 0.5$. At very late times, there
will remain only one domain boundary in the system separating two domains of size,  say, $\gamma L$ and $\beta L$,  such that
$\gamma + \beta = 1$.  With  both $\gamma , \beta > q$ the domain wall will perform random walk until either
of the domains shrinks to a size $qL$. (This picture is valid for   $ q < 0.5$ and otherwise the dynamics will be
simple model I type). Let us suppose that the domain with 
initial size $\beta L$   
shrinks to $qL$ in time $t_2^{open}$ such that the domain wall performs a random walk over a distance $s$  where
$\beta L - s = qL$. This gives
\[
t_2^{open}(\beta) \propto (\beta - q)^2L^2.
\]
Or, the average value of $t_2^{open}$ is given by
\[
t_{2}^{open} \propto \int _{q}^{1-q}  (\beta - q)^2L^2d\beta = \frac{(1-2q)^3 L^2}{3} .
\]
The result for the periodic boundary condition is  obtained by putting $q=p/2$ such that
\[
t_2 \propto (1-p)^3 L^2
\]
and therefore
\begin{equation}
t_{sat} = a pL + b (1-p)^3 L^2
\label{tsat}
\end{equation}
The above form is also consistent with the fact that $t_{sat} \propto L^2$
for $p = 0$ and $t_{sat} \propto L$ for $p=1$.
\begin{figure} [ht]
\begin{center}
 \rotatebox{270}{\resizebox*{6cm}{!}{\includegraphics{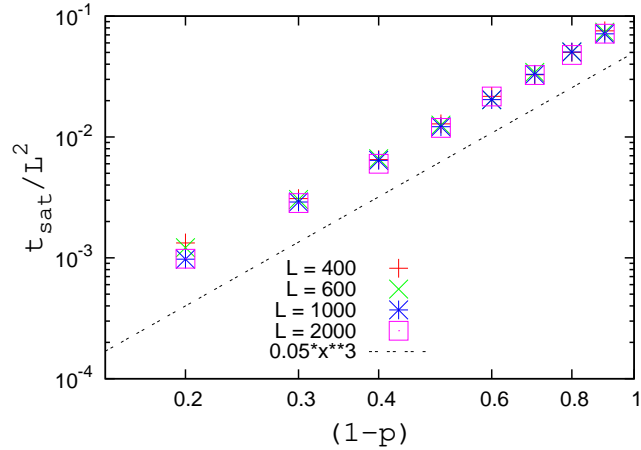}}}
\caption{ Scaled saturation time ($t_{sat}/L^2$) against $(1-p)$ for different $L$ shows collapse with $t_{sat}/L^2 \propto (1-p)^3$.}
\label{sattime}
\end{center}
\end{figure}

For large $L$, the second term in the above equation will dominate making $t_{sat} \propto (1-p)^3 L^2$.
In order to verify this, we have numerically obtained $t_{sat}$ and plotted $t_{sat}/L^2$ against $(1-p)$
for different $L$ and found a nice collapse and a fit compatible with eq (\ref{tsat}) (Fig. \ref{sattime})
with  $ a ~ \sim 1$ and $ b  ~ \sim O(10^{-2})$. We conclude therefore that in the thermodynamic limit at later times,
for any $p \neq 1$, $z=2$, i.e., the dynamics is diffusive.
This argument, in fact holds for $p \rightarrow 0$ as well showing that 
for $R$ finite, $z=2$, as discussed in the preceding section.
 
\begin{figure} [ht]
\begin{center}
 \rotatebox{0}{\resizebox*{8cm}{!}{\includegraphics{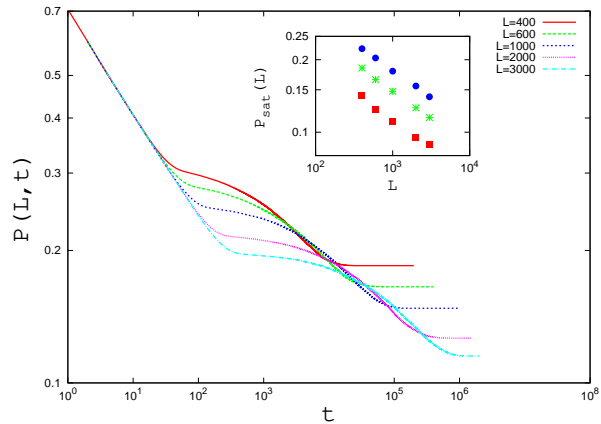}}}
\caption{ Persistence probability as a function of time for $p=0.4$ for different sizes. Inset shows
that the saturation values of the persistence probability shows a
variation $L^{-\alpha}$ for  values of $p = 0.8,0.4,0.2$ (from top to bottom) with
$\alpha \simeq 0.230$.}
\label{per}
\end{center}
\end{figure}

We have discussed so far the time dependent behaviour and exponents only. But another exponent $\alpha$
which appears at $ t \rightarrow \infty $ for the persistence probability can also be extracted here. 
The persistence probabilities show the conventional saturation at large times, with the saturation values
depending on $L$. The log-log plot of $P(L, t \to \infty)$ against $L$ shows that power law behaviour is
obeyed here with the exponent $\alpha$ once again coinciding with the model I value, $\sim 0.23$ for any value of $p \neq 0 $
(Fig. \ref{per}).

\begin{figure} [ht]
\begin{center}
 \rotatebox{0}{\resizebox*{15cm}{!}{\includegraphics{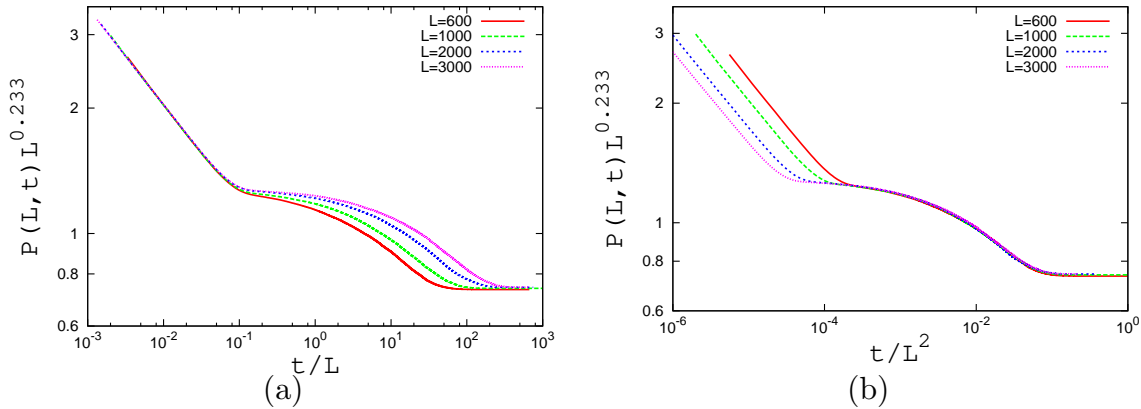}}}
\caption{ $PL^\alpha$ versus $t/L^z$ for $p=0.4$, shows a nice collapse for 
initial times up to $t_1$ using $z=1$ and $\alpha = 0.233$ (a) while using $z=2$ and the same value of $\alpha$, the data collapses over later times (b).}
\label{colpintlt}
\end{center}
\end{figure}

Having obtained $\alpha$, we  use eq (\ref{alpha}) with trial values of $z$ to obtain a collapse of
the data $PL^\alpha$ versus $t/L^z$ for any value of nonzero $p < 1$. As expected, an unique value of $z$ does not exist for
which the data will collapse over all $t/L^z$. However, we find that using $z=1$, one has a nice collapse for 
initial times up to $t_1$ while with $z=2$, the data collapses over later times (Fig. \ref{colpintlt}). The significance of the result is, an unique value of $\alpha$ is good for collapse for both time regimes. However, it is not possible
to extract any value of $\theta$ for later times as $\theta$ is extracted 
from eq (\ref{alpha}) in the limit  $t/L^z < 1$ only. 

\subsection{Discussions on the results}

At this juncture, several comments and discussions are necessary. We have obtained a crossover behaviour
in this model where an initial ballistic behaviour for macroscopic time scales is followed by a diffusive 
late time behaviour. 
However, the diffusive behaviour at later times is not apparent in the
simple log-log plots of the variables and can be extracted only from the
study of the total time to equilibriate. This is due to the fact that the 
initial ballistic dynamics leaves the system
into a non-typical configuration which is evidently far from those on 
diffusion paths. In fact in the  diffusive regime, the coarsening 
process hardly continues in terms of domain growth as only few domain walls 
remain at $t> t_1$.

A consequence of this is evident in the behaviour of the persistence at later times.  One may expect that 
the persistence exponent 3/8 may be obtained at very late times as here one has
independent random walkers, few in number, which annihilate each other 
as they meet much like in a reaction diffusion process. However, such an exponent 
is not observed from the data (Fig. \ref{percolp}).  
 Although with $z=2$ we can obtain a collapse at later times, 
it is not possible to obtain a value of $\theta$. 
Since persistence is a non-Markovian phenomena and it depends on the history, the exponent may not be apparent even 
if the phenomena is reaction diffusion like. Therefore
to analyse the dynamical 
scenario further, we study the persistence in a different way.
In order to study the persistence dynamics beyond $t=t_1$, 
we reset the zero of time at $t=t_1$. 
 In case the number of domain walls left in the system at $t_1$ is of the order of the system size ($O(L)$), 
the behaviour of persistence should be as in the case of Ising model, i.e., a power
law decay with exponent 3/8.
On the other hand, if the number of independent random walkers is {\it{finite}} (i.e., vanishes in the $L \to \infty$ limit) which
can not annihilate each other, the persistence probability is
approximately
\begin{equation}
P_{rand}(t,L) = 1-ct^{1/2}/L ,
\label{single}
\end{equation}
where we have assumed that number of distinct sites visited by the walker is proportional to the distance travelled, which is   
 $O(t^{1/2})$.

\begin{figure} [ht]
\begin{center}
 \rotatebox{0}{\resizebox*{9cm}{!}{\includegraphics{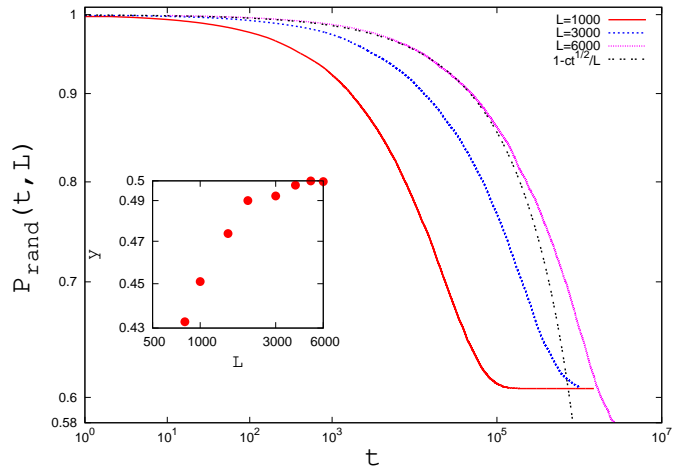}}}
\caption{ Persistence probability shows a decay as a function of time when $t_1$ is set as the initial time. 
The $L = 6000$ curve is fitted to the form $P_{rand}(t,L) = 1- ct^y/L$ with $y = 0.5$ (shown with the broken line). 
Inset shows the variation of $y$ with system size. $p=0.4$ here.}
\label{perlt}
\end{center}
\end{figure}

 We find that in the present case, resetting the zero of time at $t_1$, the persistence probability
shows a decay before attaining a constant value.
The decay for a large initial time interval can be fitted to a form $\tilde P(t) = 1- c t^y$
where the exponent $y$ increases with $L$ and 
clearly tends to saturate at 0.5 as the system size is increased. This shows that the persistence 
probability is identical to (\ref{single}) in form (Fig. \ref{perlt}). 
This  signifies that 
at $t> t_1$, the dynamics only involves the motions of random walkers which
do {\it{not}} meet and annihilate each other for a long time and explains the fact that 
domain walls remain a constant over this interval. Only at very large times close to 
equilibriation the domain walls meet and the persistence probability
starts deviating from the behaviour given by (\ref{single}). 
Actually once one of the neighbouring domains becomes less than $pL/2$ in size, the 
random walk will cease to take place and will become ballistic, which finally 
leads to  annihilation within a very short time. Therefore although we have at later times independent walkers 
performing random walk, the power law behaviour with exponent 3/8 will never be observed (even when the origin of the time is shifted) as the annihilation 
here is not taking place as in a usual reaction diffusion system but determined 
by the model I like dynamics. It may also be noted that beyond $t=t_1$,  annihilations 
occur only when the system is very close to equilibriation unlike in 
a reaction diffusion system where annihilations occur 
over all time scales.

The reason why a single value of $\alpha$ is valid
for both $t>t_1$ and $ t<t_1$ is also clear from the above study. We expect that at $t=t_1$, the number of persistent sites $\propto L^{- \alpha}$ with 
the value of $\alpha \simeq 0.235$ as in model I. The additional number of sites which become non-persistent beyond $t_1$ is proportional to $(t-t_1)^{y}/L$ and therefore at $t=t_{sat}$ expected number of persistent site is 
\[
 c_1 L^{- \alpha} - c_2 (t_{sat}-t_1)^{y}/L = c_1 L^{- \alpha} - c_2 t_2^{y}/L~,
\]
where $c_1,c_2$ are  proportionality constants. Since in the thermodynamic limit $y\rightarrow 1/2$ and $t_2 \propto L^2$, the number of persistence sites remains $\propto L^{-\alpha}$. 
Here we have assumed $c_2$ to be independent of $L$, the assumption is justified by the result.
 
\section{The case with quenched randomness}

In this section, we briefly report the behaviour of the system
when each spin is assigned a value of $p~(0<p \leq 1)$ randomly from a uniform
distribution. The randomness is quenched as the value of
$p$ assumed by a spin is fixed for all times. 

Here we note that the equilibrium behaviour, all spins up or down is
once again achieved in the system. However the time to reach equilibrium values are  larger
than the $p=1$ case. 

\begin{figure} [ht]
\begin{center}
 \rotatebox{0}{\resizebox*{10cm}{!}{\includegraphics{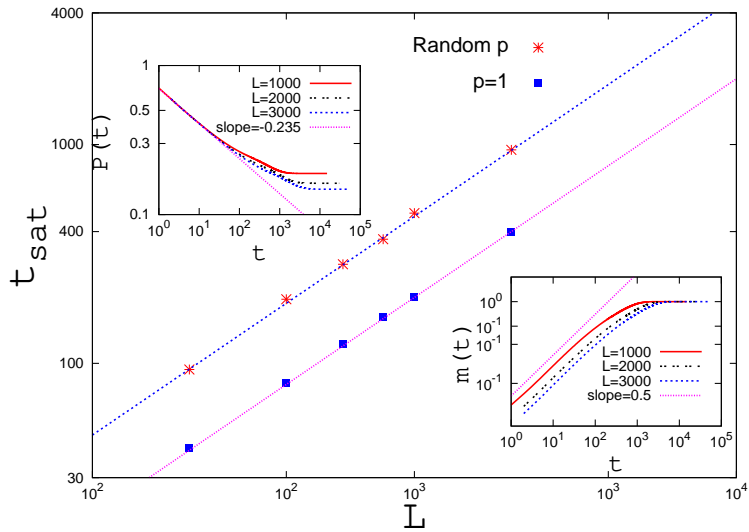}}}
\caption{ Saturation time $(t_{sat})$ against system size $L$ shows $z=1$. Inset on the top left shows the  persistence probability $P(t)$ with time which follows a power law decay with exponent $\sim 0.235$ initially.
The other inset on the bottom right shows the growth of magnetisation $m(t)$ with time where the initial variation is like $m(t) \sim t^{1/2}$.}
\label{inset3}
\end{center}
\end{figure}

The entire dynamics of the system, once again, can be
regarded as  walks performed  by the domain walls.
For $p=1$ for all sites, the walks are ballistic with the tendency
of a domain wall being to move towards its nearest one.
For $0< p \leq 1$ but same for all sites, as discussed in the previous section,
the walk is either ballistic (at initial times) or diffusive (at later times) 
but identical for all the walkers. 
 When $p$ is different for each site, one expects that 
when a site with a relatively large $p$ is hit, the corresponding  
domain wall will move towards its nearest domain wall  while when a site with relatively small $p$ 
is hit, the dynamics of the domain wall will be  diffusive.

It has been previously noted  that
model I with noise (of a different kind) 
which induces similar mixture of diffusive and ballistic
motions shows an overall ballistic behaviour (for finite noise) with the value of the 
dynamic exponent equal to  unity \cite{psnew}.  
In the present model with quenched randomness also,
we find, by analyzing the saturation times that $z=1$. However, the variation of the magnetisation, domain walls and persistence show 
power law scalings with exponents  corresponding to  model I only for an initial range of time (Fig \ref{inset3}).

\section{Summary and concluding remarks}

In summary, we have  proposed a model in which a cutoff is introduced in
the size of the neighbouring domains as  sensed  by the spins. The cutoff $R$ is expressed in terms of a parameter $p$.
 At $p \rightarrow 0$ (finite $R$) and $p=1$ 
it shows pure diffusive and ballistic behavior respectively. In the uniform case where $p$ is same for all spins, a ballistic to 
diffusive crossover occurs in time for any nonzero $ p \neq 1$. 
Usually in a crossover phenomenon, where a power law behaviour occurs with two 
different exponents, the crossover is evident from a  simple log-log plot.
In this case, however, 
the crossover phenomena  is not apparent 
as a change in exponents in simple log-log plots does not appear.
The crossover occurs between two different types of phenomena, the first
is pure coarsening in which domain walls prefer to move towards their nearest
neighbours as in model I and one gets the expected power law behaviour. At $t_1$, 
as mentioned before, some special configurations are generated and therefore the second phenomena involves pure diffusion of a few  domain walls (density of domain walls 
going to zero in the thermodynamic limit) which remain non-interacting
up to  large times. Naturally, the only dynamic exponent in the diffusive 
regime 
is the diffusion exponent $z=2$ which  is {\it {distinct}} from the growth 
exponent $z=1$. So the two dynamic exponents not only differ
in magnitude, they are connected to distinct phenomena as well.  
This
crossover behaviour is therefore a striking feature for the model.
For $R$ finite ($p \to 0$), there is no crossover effect, as the time 
$t_1$ is too small to generate these special configurations and usual 
reaction diffusion type of behaviour prevails.

Persistence probability, in whichever
way one sets the zero of time, does not show  any power law behaviour in the 
second time regime.
 At the same time, a single value of $\alpha$ is required for the
collapse in the two regimes.
 
Another point of interest is that while $z=2$ is expected for nonzero $p\neq 1$ values at later times, 
the behaviour of the total time to equilibriate as a function of $p$ is not obvious. Our
calculation shows that it is proportional to $(1-p)^3$, which is another important result 
of the present work. 


We also found that making $p$ a quenched random variable taken from an uniform
distribution,  one gets back
model I like behaviour to a large extent. 
However, choosing a different distribution might lead to different results. 
The fact that the model has different 
behaviour with uniform  $p$ and with quenched random
value of $p$ is reminiscent of the different behaviour observed in agent based models 
with savings in econophysics \cite{ccm}.


\vskip 0.5cm
Acknowledgments: Financial support from DST project SR/S2/CMP-56/2007
is  acknowledged. SB acknowledges inspiring discussions with D. Dhar.
Partial computational help has been provided by UPE project.

\end{document}